\documentclass[12pt]{article}
\usepackage{amsmath}
\usepackage{amssymb}
\usepackage{latexsym}
\usepackage{enumerate}
\usepackage{bbm}
\usepackage{amsthm}
\usepackage{hyperref}
\usepackage{graphicx}
\usepackage{dcolumn}
\usepackage{bm}
\usepackage[title]{appendix}
\newcommand{\pound}{\emph{\textsterling}}
\newcommand{\be}{\begin{equation}}
\newcommand{\ee}{\end{equation}}
\newcommand{\ben}{\begin{equation*}}
\newcommand{\een}{\end{equation*}}
\newcommand{\bea}{\begin{eqnarray}}
\newcommand{\eea}{\end{eqnarray}}
\newcommand{\bean}{\begin{eqnarray*}}
\newcommand{\eean}{\end{eqnarray*}}

\newcommand{\bsub}{\begin{subequations}}
\newcommand{\esub}{\end{subequations}}

\newcommand{\disfrac}[1][2]{\displaystyle\frac}

\newcommand{\ima}{\mathbbmtt{i}}
\newcommand{\non}{\nonumber}
\newcommand{\bbar}{\overline}

\date{}

\begin{document}

\title{{\bf Canonical Quantization of the BTZ Black Hole using Noether Symmetries}}
\vspace{1cm}
\author{\textbf{T. Christodoulakis}\thanks{tchris@phys.uoa.gr}\,, \textbf{N. Dimakis}\thanks{nsdimakis@gmail.com}\,,
\textbf{Petros A. Terzis}\thanks{pterzis@phys.uoa.gr}\\
{\it Nuclear and Particle Physics Section, Physics Department,}\\{\it University of Athens, GR 157--71 Athens}\\
\textbf{G. Doulis}\thanks{georgios.doulis@aei.mpg.de}\\
{\it Max Planck Institute for Gravitational Physics,}\\
{\it Am M\"{u}hlenberg 1, 14476 Potsdam, Germany}}
\date{}
\maketitle
\begin{center}
\textit{}
\end{center}
\vspace{-1cm}

\begin{abstract}
The well-known BTZ black hole solution of (2+1) Einstein's gravity,
in the presence of a cosmological constant, is treated both at the
classical and quantum level. Classically, the imposition of the two
manifest local Killing fields of the BTZ geometry at the level of
the full action results in a mini-superspace constraint action with
the radial coordinate playing the role of the independent dynamical
variable. The Noether symmetries of this reduced action are then
shown to completely determine the classical solution space, without
any further need to solve the dynamical equations of motion. At a
quantum mechanical level, all the admissible sets of the quantum
counterparts of the generators of the above mentioned symmetries are
utilized as supplementary conditions acting on the wave-function.
These additional restrictions, in conjunction with the
Wheeler-DeWitt equation, help to determine (up to constants) the
wave-function which is then treated semiclassically, in the sense of
Bohm. The ensuing space-times are, either identical to the classical
geometry, thus exhibiting a good correlation of the corresponding
quantization to the classical theory, or are less symmetric but
exhibit no Killing or event horizon and no curvature singularity,
thus indicating a softening of the classical conical singularity of
the BTZ geometry.

\end{abstract}

\numberwithin{equation}{section}

%%%%%%%%%%%%%%%%%%%%%%%%%%%%%%%%%%%%%%%%%%%%%%%%%%%%%%%%%%%%%%%%%%%%%%%%%%%%%%%%%%%%%%%%

\pagebreak
\section{Introduction}
\label{intro}

Due to its simplicity, three dimensional Einstein gravity is widely considered as an interesting model to explore many aspects
of General Relativity. At the classical level, it is known that any vacuum solution of Einstein's equations represents a flat
spacetime (\cite{Weinberg}, \cite{Deser1}), while the existence of a cosmological constant (again in the absence of matter)
leads to other maximally symmetric solutions, i.e. de Sitter/anti-de Sitter manifolds \cite{Deser2}. However, the study of
three dimensional geometry proved to be highly non trivial with the discovery of the BTZ black hole \cite{BTZ}. The latter
emerges in the case of pure gravity and under the presence of a negative cosmological constant $-l^2$, i.e.  the action
describing the system assumes the form
\be \label{action0}
\mathcal{A}=\int\!\!\sqrt{-g}(R+2l^2)d^3x.
\ee
Since then many aspects of the properties of the BTZ spacetime have been explored \cite{BTZ1}-\cite{Carlip3}, but it is its
canonical quantum description that motivates the present work.

Other methods have also been presented in the literature, regarding the quantization of 2+1 geometries. These, mainly focus
in solving the constraints and using them to derive a Hamiltonian description in a reduced form that results in a system of
finite degrees of freedom \cite{Witten}-\cite{Carlip2}. In this paper we study the classical and quantum description of the
BTZ geometry in a different perspective: At the classical level we are led to a finite dimensional system by imposing on a
general three dimensional line element the two manifest local isometries of the BTZ black hole ($\partial_t$ and $\partial_\phi$),
and then inserting the resulting reduced metric in the action \eqref{action0}; the result is a mini-superspace model in which
a 2+1 decomposition in the direction of the radial component $r$ is considered. This method was firstly exhibited in the case
of four dimensional static, spherically symmetric space-times in \cite{Cav1} and \cite{Cav2}. A study of the Noether symmetries
of the reduced Lagrangian leads to enough integrals of motion so that the BTZ metric is acquired without any need to solve
the dynamical equations of motion and thus identify parts of the anti-de Sitter maximal manifold. At the quantum level, we
proceed with the canonical quantization of the model; except of the quantum quadratic constraint we use the quantum analogues
of the existing Noether symmetries for the reduced system by imposing them on the wave function as eigen-operators.

Due to the reparametrization invariance under arbitrary changes of the radial coordinate ($r=f(\tilde{r})$) there is a
non-trivial problem for finding the maximum number of conserved quantities: In the theory of regular systems the infinitesimal
criterion for the determination of Noether symmetries reads $\pound_\xi L = \frac{dF}{dt}$ where $L$ is the Lagrangian and
$F$ an arbitrary function of the configuration space variables and the dynamical parameter \cite{Olver1}. However, since
mini-superspace Lagrangians are singular in nature, one has to modify the aforementioned criterion in order to acquire all
the possible Noether symmetries generated by the configuration space vector $\xi$ \cite{tchris2} (for specific examples see
\cite{tchris1}-\cite{tchris4}). In short, the required change is that the action of the generator $\xi$ on the Lagrangian
does not have to be strictly zero (or equal to a total derivative), but it suffices to be equal to a multiple of the constraint. Thus, for a mini-superspace
Lagrangian of the form
\be
L=\frac{1}{2N}\, G_{\alpha\beta}(q)\, q^\alpha q^\beta - N\, V(q),
\ee
the criterion reads $\pound_\xi L = \omega(q) \frac{\partial L}{\partial N}$, with $\omega(q)$ allowed to be any arbitrary
function. In \cite{tchris2} it was proven that the $\xi$'s that satisfy this condition are given by
\be \label{crit}
\pound_\xi (V G_{\alpha\beta}) =0,
\ee
i.e. they are Killing fields of the scaled mini-supermetric $\bbar{G}_{\alpha\beta}:=V G_{\alpha\beta}$. More over one can
also use the homothecy of the scaled mini-superspace ($\pound_{h} (V G_{\alpha\beta}) =V G_{\alpha\beta}$) to construct a
rehonomic integral of motion.

With the help of Dirac's theory for constrained systems \cite{Dirac1}-\cite{Sunder}, one can be led to a Hamiltonian description
and proceed with the quantization of the particular system. In phase space, the Noether charges defined by the $\xi$'s obtained
by \eqref{crit}, become quantities linear in the momenta. In the usual scenario of quantum cosmology one proceeds with the
quantization of the system by demanding that the action of the constraint operators being zero on the wave function. But, in
the case where Noether charges are present, one can use them to define linear Hermitian eigen-operators that can be enforced
as supplementary conditions on the wave function. Thus, simplifying the procedure and also uniquely (up to a constant phase)
defining the wave function.

The structure of the paper is as follows: In section \ref{classical} we construct the corresponding mini-superspace action,
calculate the symmetry generators and derive the classical BTZ solution with the help of the integrals of motion. In the next
section we proceed with the canonical quantization of the system and we derive a wave function for each case of possible sets
of observables. In section \ref{semicl}, we use Bohm's  semiclassical analysis \cite{Bohm}, applied on the previously found
wave functions and derive the corresponding semiclassical space-time manifolds. Finally, we conclude our analysis, summing
our results in the discussion.

\section{Classical treatment}
\label{classical}

\subsection{Derivation of the general form of the line element}
\label{sec:line_element}

We require that the metric of a $(2+1)$-dimensional spacetime with an assigned
coordinate system $(t,r,\phi)$ admits the following Killing vector fields:
\[
X_1 = \frac{\partial}{\partial t} \quad \text{and} \quad X_2= \frac{\partial}{\partial \phi}.
\]
The ensuing stationary and axisymmetric line element is of the generic form
\begin{equation}
 \label{gen_metric}
  ds^2 = g_{ij}(r)\, dx^i\,dx^j,
\end{equation}
where $i,j = 0,1,2$ and $\{x^0, x^1, x^2\} = \{t, r, \phi\}$. In order to further simplify the line element
\eqref{gen_metric}, we exploit the remaining freedom in changing coordinates in a way that does not introduce
$t$, $\phi$ in the metric components. In this spirit, inserting the following coordinate transformation
\begin{align*}
 t & \longmapsto \tilde{t} = t - \int\!\! \frac{g_{01}\,g_{22}-g_{02}\, g_{12}}{g_{02}^{2}-g_{00}\, g_{22}}\, dr, \\
 r & \longmapsto \tilde{r}=r,\\
 \phi &\longmapsto \tilde{\phi} = \phi - \int\!\! \frac{g_{00}\, g_{12}-g_{01}\, g_{02}}{g_{02}^{2}-g_{00}\, g_{22}}\, dr,
\end{align*}
(where it is assumed that $g_{02}(r)^2-g_{00}(r)\,g_{22}(r)\neq 0$), the finally reduced metric inferred from
\eqref{gen_metric} assumes the form
\begin{equation}
 \label{spec_metric}
  \tilde{g}_{ij} =
  \begin{pmatrix}
   g_{00}(r) & 0 & g_{02}(r) \\ \\
   0 & \tilde{g}_{11}(r) & 0 \\ \\
   g_{02}(r) & 0 & g_{22}(r)
  \end{pmatrix} ,
\end{equation}
with $\tilde{g}_{11}(r) = g_{11} + \disfrac{g_{00}\, g_{12}^{2}+ g_{22}\, g_{01}^{2} -
2\,g_{01}\,g_{02}\,g_{12}}{g_{02}^{2}-g_{00}\, g_{22}}$. Note that the metric \eqref{spec_metric}
guarantees that the spacetime described by it is invariant under simultaneous reflections
of the time and angular coordinate, i.e. $(\tilde{t}, \tilde{\phi}) \mapsto (-\tilde{t},
-\tilde{\phi})$.

In order to bring \eqref{spec_metric} into a form suitable for our purposes, we choose
\[
 g_{00}(r) = -a(r)^2,\quad g_{02}(r) = c(r),\quad g_{22}(r) = b(r)^2, \quad \tilde{g}_{11}(r) =\
 \frac{n(r)^2}{4\Lambda^2\left(a(r)^2\,b(r)^2+c(r)^2\right)},
\]
where $n(r)$ stands for $n(r)^2 = 4\Lambda^2 \left(g_{11} (g_{02}^{2}-g_{00}\, g_{22}) +
g_{00}\, g_{12}^{2}+ g_{22}\, g_{01}^{2}-2\,g_{01}\,g_{02}\,g_{12}\right)$ and $\Lambda$
is the cosmological constant. In this parametrization the line element \eqref{gen_metric} in the coordinate system $(\tilde{t},r,\tilde{\phi})$ assumes the form
\[
 ds^2 = -a(r)^2d\tilde{t}^2 + \frac{n(r)^2}{4\Lambda^2\left(a(r)^2\,b(r)^2+c(r)^2\right)}\,dr^2 +\
 2c(r)d\tilde{t}\,d\tilde{\phi} + b(r)^2d\tilde{\phi}^2.
\]
The above parametrisation has been chosen for several reasons: it simplifies considerable
the canonical formulation and makes the potential of the ensuing quadratic constraint constant with respect to $a(r)$, $b(r)$ and $c(r)$.
An immediate implication of the latter is that, as rigorously shown in \cite{tchris1} and easily seen from \eqref{crit},
the conformal Killing fields that generate the Noether symmetries reduce to Killing fields.

\subsection{Lagrangian and Noether (conditional) symmetries}
\label{sec:lagrangian}

According to the discussion in sec.~\ref{sec:line_element}, the stationary, axisymmetric
line element taken as the starting point in the present work is of the form
\begin{equation}
 \label{ds_BTZ}
  ds^2 = -a(r)^2dt^2 + \frac{n(r)^2}{4\Lambda^2\left(a(r)^2\,b(r)^2 + c(r)^2\right)}\,dr^2 +\
  2c(r)dt\,d\phi + b(r)^2d\phi^2,
\end{equation}
where the $\tilde{}$ were dropped for the sake of simplicity. Since we are working in a $(2+1)$
decomposition along the $r$ coordinate \cite{Cav1}, where space-time is foliated by $r$ (instead of $t$) hypersurfaces,
the field $n(r)$ plays the role of the $r$-lapse function and the
fields $a(r), b(r), c(r)$ are the $r$-dynamical variables. Notice that the $r$-lapse function is already
reparametrised, i.e. $\tilde{g}_{11}(r) = n(r)/2 \Lambda (a^2 b^2+c^2)^{1/2}$, in a way that makes the potential
of the quadratic constraint constant with respect to the dependent dynamical variables.

The action for $(2+1)$-dimensional geometries with a general cosmological constant is \eqref{action0}. For simplicity
we consider $\Lambda= l^2$ (which means that in our analysis the BTZ black hole emerges for $\Lambda>0$),
thus the aforementioned action takes the form
\begin{equation}
 \label{action}
  \mathcal{A} = \int \! \sqrt{-g}\,(R + 2 \Lambda)\, d^3 x,
\end{equation}
with $\Lambda > 0$. The action \eqref{action} applied to the geometries \eqref{ds_BTZ} reduces
to
\[
\mathcal{A} = \int \! L(a, b, c, a', b', c', n) \, dr
\]
with Lagrangian
\begin{equation}
 \label{Lag}
  L = \frac{\Lambda}{n}\,(4\,a\,b\,a'\,b' + c'^2) + n,
\end{equation}
where the $'$ denotes differentiation with respect to the radial coordinate $r$. The Lagrangian
\eqref{Lag} belongs to a particular form of singular Lagrangians: $L = \frac{1}{2\, n}\, G_{\mu\nu} \,
{q'}^\mu\,{q'}^\nu + n\, V(q)$ with ${q}^\mu = (a, b, c)$, a mini-supermetric of the form
\begin{equation}
 \label{supermetric}
  G_{\mu\nu} = 2\Lambda
  \begin{pmatrix}
   0      & 2\,a\,b & 0\\
   2\,a\,b& 0       & 0\\
   0      & 0       & 1
  \end{pmatrix}
\end{equation}
and a constant (as desired) potential $V(q) = 1$. It can be easily shown that the Euler-Lagrange
equations of \eqref{Lag} are identical to Einstein's field equations $R_{\mu\nu} - \frac{1}{2} g_{\mu\nu} R =
\Lambda g_{\mu\nu}$, resulting from the extremization of the action \eqref{action} and evaluated on the metric \eqref{ds_BTZ}.

Let us now turn attention to the conditional symmetries of \eqref{Lag}. It has been shown in \cite{tchris1}
that, in the specific lapse reparametrisation we employ in order to force the potential to depend only on the lapse,
the generators of the conditional symmetries are the Killing vector fields of the mini-supermetric
\eqref{supermetric}.This metric describes a flat, Lorentzian manifold, thus admitting a six-dimensional isometry
group of motions.  By a straightforward calculation it can be confirmed that the infinitesimal
condition $\mathcal{L}_\xi G_{\mu\nu} = 0$ is satisfied by the following six generators:
\begin{equation}
\begin{split}
 \label{kill_vect}
 &\xi_1 = \partial_c, \quad \xi_2 = \frac{c}{2b}\,\partial_b - \frac{a^2}{2}\,\partial_c, \quad \xi_3 = -\frac{c}{a}\,\partial_a + b^2\,\partial_c, \\
 &\xi_4 = \frac{1}{2b}\,\partial_b, \quad \xi_5 = -a\,\partial_a + b\,\partial_b, \quad \xi_6 = \frac{1}{a}\,\partial_a.
\end{split}
\end{equation}
The above fields satisfy a Lie algebra $[\xi_K, \xi_L] = C^M_{KL} \xi_M$ of which we present only the
non-zero structure constants:
\begin{align}
 \label{str_const}
  &C^4_{12} = C^6_{31} = C^1_{26} = C^1_{43} = -C^4_{21} = -C^6_{13} = -C^1_{62} = -C^1_{34} = 1,\non \\
  &C^2_{25} = C^3_{53} = C^4_{45} = C^6_{56} = -C^2_{52} = -C^3_{35} = -C^4_{54} = -C^6_{65} = 2, \\
  &C^5_{32} = -C^5_{23} = \frac{1}{2}.\non
\end{align}
In addition, \eqref{supermetric} admits a homothetic vector field that satisfies the condition
$\mathcal{L}_h G_{\mu\nu} = G_{\mu\nu}$:
\begin{equation}
 \label{homo_vect}
 h = \frac{b}{2}\,\partial_b + \frac{c}{2}\,\partial_c.
\end{equation}

\subsection{Hamiltonian formulation and the solution space}
\label{sec:hamiltonian}

We now develop the Hamiltonian formulation of the geometries \eqref{ds_BTZ} and derive the classical solution
space by algebraic means. Following \cite{Dirac, Sunder} we first define the $r$-conjugate momenta
\begin{subequations}
 \label{momenta}
  \begin{align}
   &\pi_n = \frac{\partial L}{\partial n'} = 0, \label{momentum_n} \\
   &\pi_a = \frac{\partial L}{\partial a'} = \frac{4\, \Lambda\, a\, b\, b'}{n}, \label{momentum_a} \\
   &\pi_b = \frac{\partial L}{\partial b'} = \frac{4\, \Lambda\, a\, b\, a'}{n}, \label{momentum_b} \\
   &\pi_c = \frac{\partial L}{\partial c'} = \frac{2\, \Lambda\,  c'}{n}, \label{momentum_c}
  \end{align}
\end{subequations}
where $\pi_n$ is a first class primary constraint. Invoking the Legendre transformation for \eqref{Lag},
we arrive at the Hamiltonian
\[
 H = n\, \mathcal{H} + u_n \pi_n
\]
where the functions $u_n$ are Lagrangian multipliers and
\begin{equation}
 \label{quadr_constr}
  \mathcal{H} = \frac{\pi_a\, \pi_b}{4\, \Lambda\, a\, b}+\frac{\pi_c^2}{4 \Lambda } - 1,
\end{equation}
which, in view of the need for preservation of the primary constraint $\pi_n$ during the evolution,
i.e. $\pi_n'=\{\pi_n, H\} \approx 0$, leads to the first class secondary constraint $\mathcal{H} \approx 0$.

According to \cite{tchris1}, each one of the Killing vector fields contracted with the non-vanishing
conjugate momenta \eqref{momenta}, i.e. $Q_I:=\xi_I^\mu \pi_\mu$, corresponds, on the phase space, to
a linear integral of motion:
\begin{equation}
 \label{int_motion}
  \begin{aligned}
   &Q_1 = \pi_c, \quad Q_2 = \frac{c}{2\,b}\,\pi_b - \frac{a^2}{2}\,\pi_c, \quad Q_3 = -\frac{c}{a}\,\pi_a + b^2\,\pi_c, \\
   &Q_4=\frac{1}{2\,b}\,\pi_b, \quad Q_5=-a\,\pi_a + b\,\pi_b, \quad Q_6 = \frac{1}{a}\,\pi_a.
  \end{aligned}
\end{equation}
The phase space quantities \eqref{int_motion} form a Poisson algebra $\{Q_K, Q_L\} = C^M_{KL} Q_M$
with structure constants given by \eqref{str_const}. The Poisson brackets of \eqref{int_motion} with
the Hamiltonian $H$ are strongly vanishing, i.e. $\{Q_I, H\} = 0$, guaranteeing that the six quantities
\eqref{int_motion} are constants of motion:
\begin{equation}
 \label{const_motion}
  Q_I = \kappa_I.
\end{equation}
In addition, from the quadratic constraint \eqref{quadr_constr} and \eqref{int_motion}, one can easily
read off the first Casimir invariant of the algebra formed by \eqref{int_motion}, namely
\begin{equation}
 \label{casimir1}
  Q_C = \mathcal{H} + 1 = \frac{Q_1^2}{4\, \Lambda } + \frac{Q_4\, Q_6}{2\, \Lambda}.
\end{equation}
One observes that in this parametrization of the constant potential, $Q_C$ is the kinetic part of the Hamiltonian.
An appropriate combination of the phase space quantities $Q_I$ leads to the second Casimir invariant
\begin{equation}
 \label{casimir2}
  \bbar{Q}_C = 2\, Q_2 Q_6 + 2\, Q_3 Q_4 - Q_1 Q_5,
\end{equation}
which vanishes identically when the quantities $Q_I$ are expressed via \eqref{int_motion} in terms
of the conjugate momenta. Let us define now the phase space quantity corresponding to the homothetic
vector field \eqref{homo_vect}:
\begin{equation}
 \label{homo_quant}
  Q_h = \frac{b\, \pi_b}{2} + \frac{c\, \pi_c}{2},
\end{equation}
whose Poisson bracket with the Hamiltonian $H$ reads
\[
 \{Q_h, H\} = H + n \approx n,
\]
in view of the vanishing of the quadratic constraint \eqref{quadr_constr} on the constraint surface.
The latter result implies that $\frac{d Q_h}{dr} = n$, which as it has been shown in \cite{tchris2}
by integration over $r$ becomes a rheonomic integral of the form
\begin{equation}
 \label{rheon_int}
  Q_h - \int \!\! n\, dr = const.
\end{equation}

In the rest of this section, we will show that \eqref{int_motion} together with \eqref{rheon_int}
and the Casimir invariants \eqref{casimir1}-\eqref{casimir2} are enough to determine the entire
classical solution space of the geometries \eqref{ds_BTZ}. The integrals of motion \eqref{int_motion},
\eqref{rheon_int}, and \eqref{casimir1}-\eqref{casimir2} become constants on the solution space;
thus, the following relations readily follow
\begin{subequations} \label{int_motion_sol}
 \begin{align}
   & Q_I = \kappa_I, \quad I = 1,\ldots,6, \label{Q_I_sol} \\
   & Q_h - \int\!\! n\, dr = c_h, \label{Q_h_sol} \\
   & 4\, \Lambda = \kappa_1^2 + 2\, \kappa_4\, \kappa_6, \label{Q_C1_sol} \\
   & 0 = 2\, \kappa_2 \kappa_6 + 2\, \kappa_3 \kappa_4 - \kappa_1 \kappa_5 , \label{Q_C2_sol}
 \end{align}
\end{subequations}
where $\kappa_I, c_h$ are constants; in the derivation of the last relations it was taken into
account that $\mathcal{H} = 0$ on the solution space and that $\bbar{Q}_C = 0$. Let us first solve
algebraically the five first ($I = 1,\ldots,5$) equations of \eqref{Q_I_sol} together with \eqref{Q_h_sol}
for $a,\, a',\, c,\, c',\, n,$ and $\int\!\! n\, dr$:
\begin{subequations}
 \label{algebr_sol}
  \begin{align}
   & a = \sqrt{\frac{2\,\kappa_2 (2\, \kappa_4 b^2 - \kappa_5)}{\kappa_1 \kappa_5 - 2\, \kappa_3 \kappa_4}}\, , \label{algebr_sol_a} \\
   & a' = 2\,\kappa_4 b\, b'\sqrt{\frac{2\,\kappa_2}{(2\, \kappa_4 b^2 - \kappa_5)(\kappa_1 \kappa_5 - 2\, \kappa_3 \kappa_4)}}\, , \label{algebr_sol_a'} \\
   & c = \frac{2\, \kappa_2 \left(\kappa_3 - \kappa_1 b^2\right)}{2\, \kappa_3 \kappa_4 - \kappa_1 \kappa_5}\, , \label{algebr_sol_c} \\
   & c' = \frac{4\, \kappa_1 \kappa_2\, b\, b'}{\kappa_1 \kappa_5 - 2\, \kappa_3 \kappa_4}\, , \label{algebr_sol_c'} \\
   & \int\!\! n\, dr = \frac{b^2 \left(\kappa_1^2 \kappa_2 + \kappa_1 \kappa_4 \kappa_5 - 2\, \kappa_3 \kappa_4^2\right) -\
     \kappa_1 (\kappa_2 \kappa_3 + \kappa_5 \kappa_7) + 2\, \kappa_3 \kappa_4 \kappa_7}{\kappa_1 \kappa_5 - 2\, \kappa_3 \kappa_4}\, , \label{algebr_sol_int_n} \\
   & n = \frac{8\, \kappa_2\, \Lambda\, b\, b'}{\kappa_1 \kappa_5 - 2\, \kappa_3\kappa_4}\, , \label{algebr_sol_n}
  \end{align}
\end{subequations}
where $b$ remains an arbitrary function of $r$. It can be easily checked that the consistency conditions for the fields $a$
and $c$ ($a'=\frac{da}{dr}$ and $c'=\frac{dc}{dr}$) are identically satisfied in view of \eqref{algebr_sol_a}-\eqref{algebr_sol_a'}
and \eqref{algebr_sol_c}-\eqref{algebr_sol_c'}, respectively. The consistency condition for $n$ ($n=\frac{d}{dr}\int ndr$),
on the other hand, requires
\[
 -\kappa_2 (\kappa_1^2 - 4\, \Lambda) + \kappa_4 (2\, \kappa_3 \kappa_4 - \kappa_1 \kappa_5) = 0,
\]
which, in view of \eqref{Q_C1_sol} and \eqref{Q_C2_sol}, is identically satisfied. To complete our consistency
check, we substitute \eqref{algebr_sol} into the equation $Q_6 = \kappa_6$ that has not been used in
the above derivation; the result reads
\[
 2\, \kappa_2 \kappa_6 + 2\, \kappa_3 \kappa_4 - \kappa_1 \kappa_5 = 0,
\]
which is identically satisfied by virtue of \eqref{Q_C2_sol}. Now, it is an easy task to verify that
\eqref{algebr_sol} together with \eqref{Q_C1_sol} and \eqref{Q_C2_sol} solve Einstein's field equations
$R_{\mu\nu} - \frac{1}{2} g_{\mu\nu} R = \Lambda g_{\mu\nu}$.

It is obvious from the discussion above that on the solution space there are six constants available
and two equations, i.e. \eqref{Q_C1_sol}-\eqref{Q_C2_sol}, constraining them; thus, we can specify
freely four of them. The following choice of the constants $\kappa_I$ (that respects both conditions
\eqref{Q_C1_sol}-\eqref{Q_C2_sol})
\begin{equation}
 \label{kappas}
  \begin{aligned}
   & \kappa_1 = d, \quad \kappa_2 = \frac{1}{8}(-d^2\, J + 4\, d\, M - 4\, J\, \Lambda), \quad \kappa_3 = J, \\
   & \kappa_4 = \frac{1}{4} (d^2 - 4\, \Lambda), \quad \kappa_5 = d\, J - 2\, M, \quad \kappa_6 = -2,
  \end{aligned}
\end{equation}
accompanied by a re-definition of the angular coordinate $\phi \mapsto \bar{\phi} = \phi - \frac{d}{2}\,t$
brings the line element \eqref{ds_BTZ} into the form
\begin{equation}
 \label{BTZ_metric_b}
  ds^2 = (M - \Lambda\, b^2)dt^2 + \frac{4\, b^2 b'^2}{ J^2 - 4\, M\, b^2 + 4\, \Lambda\, b^4}\,dr^2 +\
  J\, dt\,d\phi + b^2d\phi^2
\end{equation}
which, when $b$ is chosen as $b(r) = r$, exactly reproduces the BTZ metric originally introduced in \cite{BTZ}:
\begin{equation}
 \label{BTZ_metric}
  ds^2 = (M - \Lambda\, r^2)dt^2 + \left(- M + \Lambda\, r^2 + \frac{J^2}{4\, r^2}\right)^{-1} dr^2 +\
  J\, dt\,d\phi + r^2d\phi^2.
\end{equation}
From \eqref{BTZ_metric} it is obvious that $d$ is a non-essential constant and can, since it is additively absorbed,
be set to zero in order to simplify the expressions of the constants $\kappa_I$. Thus, by setting $d = 0$ the constants
\eqref{kappas} considerably
simplify to
\begin{equation}
 \label{kappas_0}
  \kappa_1 = 0, \quad \kappa_2 = \frac{J\, \Lambda}{2}, \quad \kappa_3 = J, \quad
  \kappa_4 = -\Lambda, \quad \kappa_5 = - 2\, M, \quad \kappa_6 = -2.
\end{equation}
Notice that by using the above choice of constants one can bring \eqref{ds_BTZ} into the form
\eqref{BTZ_metric} directly without re-defining the $\phi$ coordinate.

%%%%%%%%%%%%%%%%%%%%%%%%%%%%%%%%%%%%%%%%%%%%%%%%%%%%%%%%%%%%%%%%%%%%%%%%%%%%%%%%%%%%%%%%

\section{Canonical quantization using Noether symmetries}

\subsection{General considerations}
\label{general_quantum}

Here, we will quantize the classical system described in sec.~\ref{classical} according to
Dirac's canonical quantization procedure \cite{Dirac} supplemented by the condition that the
wave function must be an eigenfunction of the quantum analogs of the generators of the Noether
symmetries \eqref{int_motion}. The method has been extensively described in \cite{tchris1}
and applied to the quantization of the Schwarzshild \cite{tchris1} and Reissner-Nordstr\"{o}m
\cite{tchris3} black holes. Below, we give a brief overview of the method described in \cite{tchris1}.

As usual, in the Schr\"{o}dinger representation the classical dynamical variables become
operators
\[
 \widehat{q}^\alpha := q^\alpha \quad \mathrm{and}  \quad \widehat{\pi}_\alpha := -\ima \,  \hbar \frac{\partial}{\partial q^{\alpha}},
\]
where $q^\alpha = \{a, b, c, n\}$. (In the following we assume that $\hbar = 1$.) The above
operators obey the canonical commutation relations $[\widehat{q}^\alpha, \widehat{\pi}_\beta] =
\ima\, \delta^\alpha{}_\beta$. Following Dirac's proposal \cite{Dirac} we demand the quantum
analogs of the first class constaints \eqref{momentum_n} and \eqref{quadr_constr} to annihilate
the wavefuntion, namely
\begin{align*}
 &\widehat{\pi}_n \Psi(a,b,c,n) = -\ima\, \frac{\partial}{\partial n} \Psi(a,b,c,n) = 0 \Rightarrow \Psi = \Psi(a,b,c), \\
 &\widehat{\mathcal{H}}\Psi = (\widehat{Q}_C - 1)\Psi = 0,
\end{align*}
where the former condition guarantees that the wave function is lapse-independent and the
latter is the unit eigenvalue problem for the Casimir invariant \eqref{casimir1}. The quantum
analog of the Casimir invariant $Q_C$ in the above expression is given as the most general scalar
quadratic Hermitian (under an arbitrary measure $\mu$) operator \cite{tchris1}
\begin{equation}
 \label{quadr_operator}
 \widehat{Q}_C = - \frac{1}{2\mu} \partial_\alpha (\mu \,G^{\alpha\beta} \partial_\beta),
\end{equation}
a choice by which the latter of the above conditions becomes
\begin{equation}
 \label{quadr_quant_cond}
  \widehat{\mathcal{H}}\Psi = \left(-\frac{1}{2\mu} \partial_\alpha (\mu \,G^{\alpha\beta} \partial_\beta) - 1\right)\Psi = 0.
\end{equation}
In a similar fashion, we define the quantum analogs of the integrals of motion \eqref{int_motion}
as the most general linear Hermitian (under the same measure $\mu$) operators
\begin{equation}
 \label{linear_operator}
  \widehat{Q}_I := - \frac{\ima}{2\mu} \left(\mu \, \xi_I^\alpha\,\partial_\alpha + \partial_\alpha \, \mu\, \xi_I^\alpha \right).
\end{equation}
In \cite{tchris1}, it has been proven that linear operators of the form \eqref{linear_operator}
satisfy the same algebra as their classical counterparts $Q_I$, i.e.
\begin{equation}
 \label{quant_algebra}
  [\widehat{Q}_K, \widehat{Q}_L] =\ima\, C^M_{KL} \widehat{Q}_M,
\end{equation}
where the structure constants $C^M_{KL}$ are given by \eqref{str_const}.
Now, we can form an eigenvalue problem for each one of the linear operators \eqref{linear_operator}
\begin{equation}
 \label{linear_quant_cond}
  \widehat{Q}_I\Psi = - \frac{\ima}{2\mu} \left(\mu \, \xi_I^\alpha\,\partial_\alpha +
  \partial_\alpha \, \mu\, \xi_I^\alpha \right) \Psi = \kappa_I \Psi,
\end{equation}
where $\kappa_I$ are the eigenvalues of the operators $\widehat{Q}_I$. For several reasons,
see \cite{tchris1}, a natural geometric choice of the measure reads
\begin{equation}
 \label{measure}
  \mu = \sqrt{\det G_{\alpha\beta}} = 4 \sqrt{2\, \Lambda^3}\,  a\, b.
\end{equation}

It was shown in \cite{ChrisPap} and \cite{tchris1} that \eqref{linear_quant_cond} together with the quantum
algebra of the operators $\widehat{Q}_I$ impose certain restrictions on which of the linear operators
$\widehat{Q}_I$ can be simultaneously applied on the wave function. This selection rule
is given by the integrability condition
\begin{equation}
 \label{selec_rule}
  C^M_{KL}\, \kappa_M = 0.
\end{equation}
In addition, the number of essential constants of the underlying geometry provide a lower
bound on the number of linear operators that must be used to define eigenvalue equations simultaneously on the wave
function; in our case the relevant constants involved are two: the mass $M$ and the angular
momentum $J$ of the black hole.

At this point a clarification concerning the nature of $M$, $J$ is pertinent: their origin, as explained
in \cite{BTZ1}, can be traced to identifications of parts of the maximal AdS(3) manifold. In that global
(topological) sense they are essential for characterizing the BTZ space-time. However, locally they can
be absorbed by appropriate coordinate transformations---see \cite{gop} for the infinitesimal criterion
and Appendix \ref{appA} for the actual construction of the transformation. We make this distinction since, as we
will see later, in the semiclassical approximation $M$ and $J$ become also locally essential.

In the following, we will say that the operators $\widehat{Q}_I$
satisfying simultaneously the above selection rule form an admissible subalgebra of the
full quantum algebra \eqref{quant_algebra}, otherwise we will say that they form a non-admissible
subalgebra. Let us now exemplify the use of the selection rule \eqref{selec_rule}. First,
consider the three dimensional subalgebra $\{\widehat{Q}_1, \widehat{Q}_2, \widehat{Q}_3\}$.
Observing \eqref{str_const} the condition \eqref{selec_rule} for each combination of operators
gives
\begin{equation*}
\begin{split}
 &C^M_{12}\, \kappa_M = C^4_{12}\, \kappa_4 = \kappa_4 = 0, \quad
 C^M_{13}\, \kappa_M = C^6_{13}\, \kappa_6 = -\kappa_6 = 0, \\
 &C^M_{23}\, \kappa_M = C^5_{23}\, \kappa_5 = -\frac{1}{2}\, \kappa_5 = 0;
\end{split}
\end{equation*}
thus, if one wants the subalgebra $\{\widehat{Q}_1, \widehat{Q}_2, \widehat{Q}_3\}$ to
satisfy \eqref{selec_rule} one must set $\kappa_4 = \kappa_5 = \kappa_6 = 0$, a condition
that cannot be met in view of \eqref{kappas} as $\kappa_6 = -2$. Therefore, $\{\widehat{Q}_1,
\widehat{Q}_2, \widehat{Q}_3\}$ is a non-admissible subalgebra. Next, consider the three
dimensional subalgebra $\{\widehat{Q}_1, \widehat{Q}_4, \widehat{Q}_6\}$. A similar
computation results in
\[
 C^M_{14}\, \kappa_M = 0, \quad
 C^M_{16}\, \kappa_M = 0, \quad
 C^M_{46}\, \kappa_M = 0,
\]
which are identically satisfied because of the vanishing of all the structure constants
involved. Thus, $\{\widehat{Q}_1, \widehat{Q}_4, \widehat{Q}_6\}$ is an admissible subalgebra.
Below, we list all the admissible subalgebras we are going to subsequently consider.\\
\\
{\bf Admissible subalgebras}:
\begin{itemize}
 \item Three dimensional subalgebras:
  \begin{equation}
   \label{3_subalg}
    \{\widehat{Q}_1, \widehat{Q}_4, \widehat{Q}_6\}
  \end{equation}
 \item Two dimensional subalgebras:
  \begin{subequations}
   \label{2_subalg}
    \begin{align}
     \{\widehat{Q}_1, \widehat{Q}_5\}, \label{2_subalg_a}\\
     \{\widehat{Q}_2, \widehat{Q}_4\}, \label{2_subalg_b}\\
     \{\widehat{Q}_3, \widehat{Q}_6\}. \label{2_subalg_c}
    \end{align}
  \end{subequations}
\end{itemize}

Summarising, the system of differential equations \eqref{quadr_quant_cond}, \eqref{linear_quant_cond}
will be solved (for the choice of measure \eqref{measure}) for each one of the admissible subalgebras \eqref{3_subalg}-\eqref{2_subalg_c}.

\subsection{The three dimensional subalgebra \{\texorpdfstring{$\widehat{Q}_1, \widehat{Q}_4, \widehat{Q}_6$}{Q1, Q4, Q6}\}}
\label{sec:Q1_Q4_Q6_quantum}

Let us start with the three dimensional subalgebra \eqref{3_subalg}. There are three conditional
symmetries, for each one of which an eigenvalue problem of the form \eqref{linear_quant_cond}
must be defined. For the choice of measure \eqref{measure} and the Killing vector fields
\eqref{kill_vect} these eigenvalue problems read
\begin{subequations}
 \begin{align}
  & \kappa_1 \Psi + \ima\, \partial_c \Psi = 0, \label{linear_Q1_Q4_Q6_a} \\
  & \kappa_4 \Psi + \frac{\ima}{2 b}\, \partial_b \Psi = 0, \label{linear_Q1_Q4_Q6_b} \\
  & \kappa_6 \Psi + \frac{\ima}{a}\, \partial_a \Psi = 0. \label{linear_Q1_Q4_Q6_c}
 \end{align}
\end{subequations}
In addition, we have to solve the quadratic constaint \eqref{quadr_quant_cond}
\begin{equation}
 \label{quadr_Q1_Q4_Q6}
  \frac{1}{4\, \Lambda}\, \partial_{cc}\Psi + \frac{1}{4\, a\, b\, \Lambda}\, \partial_{ab}\Psi + \Psi = 0.
\end{equation}
Integrating successively from \eqref{linear_Q1_Q4_Q6_a} to \eqref{linear_Q1_Q4_Q6_c},
one obtains the general solution
\begin{equation}
 \label{sol_Q1_Q4_Q6}
  \Psi = c_0 e^{\frac{1}{2} \ima \left(\kappa_6 a^2 + 2\, \kappa_4 b^2 + 2\, \kappa_1 c \right)},
\end{equation}
where $c_0$ is an arbitrary constant. Inserting the latter expression into \eqref{quadr_Q1_Q4_Q6}
we get $\kappa_1^2 + 2\, \kappa_4 \kappa_6 - 4\, \Lambda = 0$, which is identically satisfied
in view of \eqref{Q_C1_sol}.

\subsection{The two dimensional subalgebra \{\texorpdfstring{$\widehat{Q}_1, \widehat{Q}_5$}{Q1, Q5}\}}
\label{sec:Q1_Q5_quantum}

Next, let us consider the two dimensional subalgebra \eqref{2_subalg_a}. The two conditional
symmetries dictate the following two eigenvalue problems
\begin{subequations}
 \begin{align}
  & \kappa_1 \Psi + \ima\, \partial_c \Psi = 0, \label{linear_Q1_Q5_a} \\
  & \kappa_5 \Psi - \ima\, a\, \partial_a \Psi + \ima\, b\, \partial_b \Psi = 0, \label{linear_Q1_Q5_b}
 \end{align}
\end{subequations}
which restrict the form of the wave function to
\[
 \Psi(a,b,c) = a^{-\ima\, \kappa_5} e^{\ima\, \kappa_1 c} \psi(a\, b),
\]
where $\psi$ is an arbitrary function of $a\,b$. The Hamiltonian constraint \eqref{quadr_Q1_Q4_Q6}
further restricts the solution to its final form
\begin{equation}
 \label{sol_Q1_Q5}
  \Psi = \left(\frac{b}{a}\right)^{\ima\, \frac{\kappa_5}{2}} e^{\ima\, \kappa_1 c}
  \left[c_1\, J_{\frac{\ima\, \kappa_5}{2}}\left(-\ima\, a\, b \sqrt{\kappa_1^2 - 4\, \Lambda}\right) +
  c_2\, Y_{\frac{\ima\, \kappa_5}{2}}\left(-\ima\, a\, b \sqrt{\kappa_1^2 - 4\, \Lambda}\right)\right],
\end{equation}
where $c_1, c_2$ are arbitrary constants and $J_\nu(z)$ and $Y_\nu(z)$ are the Bessel
functions of the first and second kind, respectively.

\subsection{The two dimensional subalgebra \{\texorpdfstring{$\widehat{Q}_2, \widehat{Q}_4$}{Q2, Q4}\}}
\label{sec:Q2_Q4_quantum}

The eigenvalue problems of the form \eqref{linear_quant_cond} for the two dimensional
subalgebra \eqref{2_subalg_b} read
\begin{subequations}
 \begin{align}
  & \kappa_2 \Psi + \frac{\ima\, c}{2 b}\, \partial_b \Psi - \frac{\ima\, a^2}{2}\, \partial_c \Psi = 0, \label{linear_Q2_Q4_a} \\
  & \kappa_4 \Psi + \frac{\ima}{2 b}\, \partial_b \Psi = 0, \label{linear_Q2_Q4_b}
 \end{align}
\end{subequations}
which bring the wave function into the form
\[
 \Psi = \psi(a)\, e^{\frac{\ima\, \left(\kappa_4 a^2 b^2 + \kappa_4 c^2 - 2\, \kappa_2 c\right)}{a^2}},
\]
where $\psi$ is an arbitrary function of $a$. The Hamiltonian constraint \eqref{quadr_Q1_Q4_Q6}
specifies the function $\psi$ and leads to the general solution
\begin{equation}
 \label{sol_Q2_Q4}
  \Psi = \frac{c_0}{a}\, e^{\frac{\ima\, \left(a^4 \Lambda + \kappa_4^2 a^2 b^2 + \kappa_4^2 c^2 -
  2\, \kappa_2 \kappa_4 c + \kappa_2^2\right)}{\kappa_4 a^2}},
\end{equation}
where $c_0$ is an arbitrary constant.

\subsection{The two dimensional subalgebra \{\texorpdfstring{$\widehat{Q}_3, \widehat{Q}_6$}{Q3, Q6}\}}
\label{sec:Q3_Q6_quantum}

Finally, we will consider the two dimensional subalgebra \eqref{2_subalg_c}. The two
eigenvalue problems read
\begin{subequations}
 \begin{align}
  &  \kappa_3 \Psi + \ima\, b^2 \partial_c \Psi - \frac{\ima\, c}{a}\, \partial_a \Psi = 0, \label{linear_Q3_Q6_a} \\
  & \kappa_6 \Psi + \frac{\ima}{a}\, \partial_a \Psi = 0. \label{linear_Q3_Q6_b}
 \end{align}
\end{subequations}
The two differential equations above allow the solution
\[
 \Psi = \psi(b)\, e^{\frac{\ima\, \left(\kappa_6 a^2 b^2 + \kappa_6 c^2 + 2\, \kappa_3 c\right)}{2\, b^2}}
\]
for an arbitrary function $\psi(b)$. Taking into account the quadratic constraint \eqref{quadr_Q1_Q4_Q6}
one arrives at the general solution
\begin{equation}
 \label{sol_Q3_Q6}
  \Psi = \frac{c_0}{b}\, e^{\frac{\ima\, \left(\kappa_6^2 a^2 b^2 + 4\,  \Lambda\, b^4 + \kappa_6^2 c^2 +
   2\, \kappa_3 \kappa_6 c + \kappa_3^2 \right)}{2\, \kappa_6 b^2}},
\end{equation}
where $c_0$ is an arbitrary constant.

%%%%%%%%%%%%%%%%%%%%%%%%%%%%%%%%%%%%%%%%%%%%%%%%%%%%%%%%%%%%%%%%%%%%%%%%%%%%%%%%%%%%%%%%

\section{Semiclassical analysis}
\label{semicl}

\subsection{General considerations}
\label{sec:general_quantum}

To make a connection between the quantum solutions \eqref{sol_Q1_Q4_Q6}, \eqref{sol_Q1_Q5},
\eqref{sol_Q2_Q4}, and \eqref{sol_Q3_Q6} of the various admissible subalgebras and the
classical solution space \eqref{BTZ_metric}, we develop here a semiclassical analysis
of these quantum results in the spirit of \cite{tchris3} following the original Bohmian approximation \cite{Bohm}.

We start by defining the general form of the wave function
\begin{equation}
 \label{wave}
  \Psi(a,b,c) = \Omega(a,b,c)\, e^{\ima\, S(a,b,c)},
\end{equation}
where $\Omega(a,b,c)$ and $S(a,b,c)$ are the amplitude and the phase of the wave function,
respectively. Inserting \eqref{wave} into the Hamiltonian constraint \eqref{quadr_Q1_Q4_Q6}
and taking the imaginary part, one arrives at the continuity equation
\begin{equation}
 \label{contin_eq}
  \frac{1}{2\, \Lambda} \left(\frac{1}{2\, a\, b}\left(\partial_a S\, \partial_b \Omega +
  \partial_b S\, \partial_a \Omega + \Omega\, \partial_{ab} S \right) +
  \partial_c S\, \partial_c \Omega + \frac{\Omega}{2}\, \partial_{cc} S \right) = 0.
\end{equation}
The real part reads
\begin{equation}
 \label{hamil_jac}
  \frac{\partial_a S\, \partial_b S}{4\, \Lambda\, a\, b} + \frac{\partial_c S\, \partial_c S}{4\, \Lambda} - 1 - \mathcal{V} = 0,
\end{equation}
where we define the quantum potential (its name will be shortly justified)
\begin{equation}
 \label{quant_pot}
  \mathcal{V} = \frac{1}{4\, \Lambda\, \Omega} \left(\frac{\partial_{ab} \Omega}{a\, b} + \partial_{cc} \Omega \right).
\end{equation}
Notice that for a vanishing quantum potential \eqref{hamil_jac} resembles the Hamiltonian
constraint \eqref{quadr_constr}; in that case, a direct inspection of their elements leads
to the identification $\{\pi_a, \pi_b, \pi_c\} = \{\partial_a S, \partial_b S, \partial_c S\}$,
which when expressed, through \eqref{momentum_a}-\eqref{momentum_c}, in terms of the variables
of the configuration space provide us with the following semiclassical equations of motion
\begin{equation}
 \label{semicl_eqs}
  \begin{aligned}
   & \partial_a S - \frac{4\, a\, b\, b'\, \Lambda}{n} = 0, \\
   & \partial_b S - \frac{4\, a\, a'\, b\, \Lambda}{n} = 0, \\
   & \partial_c S - \frac{2\, c'\, \Lambda}{n} = 0.
  \end{aligned}
\end{equation}
From the discussion above one expects that when the quantum potential vanishes, $\mathcal{V} = 0$,
the semiclassical equations \eqref{semicl_eqs} must reproduce the classical solution \eqref{BTZ_metric}.
Otherwise, when $\mathcal{V} \neq 0$, the solution of the semiclassical equations must differ
from the classical one because of quantum effects introduced by the quantum potential.

\subsection{The three dimensional subalgebra \{\texorpdfstring{$\widehat{Q}_1, \widehat{Q}_4, \widehat{Q}_6$}{Q1, Q4, Q6}\}}
\label{sec:Q1_Q4_Q6_semi}

Let us start with the wave function \eqref{sol_Q1_Q4_Q6},
$\Psi = c_0\, e^{\frac{1}{2} \ima \left(\kappa_6 a^2 + 2\, \kappa_4 b^2 + 2\, \kappa_1 c \right)}$,
of the three dimensional subalgebra \eqref{3_subalg}. Inserting the numerical values \eqref{kappas_0}
of the $\kappa_I$'s corresponding to the classical solution space, the above wave function
reduces to
\[
 \Psi = c_0\, e^{\frac{1}{2} \ima \left(-2\, a^2 - 2\, b^2 \Lambda \right)}.
\]
A direct comparison of the latter with \eqref{wave} leads to the identification
\begin{equation}
 \label{ampl_phas_Q1_Q4_Q6}
  \Omega = c_0, \qquad S = \frac{1}{2} \left(-2\, a^2 - 2\, b^2 \Lambda \right),
\end{equation}
which satisfy the continuity equation \eqref{contin_eq} and lead to a vanishing quantum
potential $\mathcal{V} = 0$. Thus, it is expected that the solution to the following
semiclassical equations of motion
\begin{subequations}
 \label{eq_mot_Q1_Q4_Q6}
  \begin{align}
   & \frac{2\, a\, b\, b'}{r} - 2\, a = 0, \label{eq_mot_Q1_Q4_Q6_a} \\
   & \frac{2\, a\, a'\, b}{r} - 2\, \Lambda\, b = 0, \label{eq_mot_Q1_Q4_Q6_b} \\
   & \frac{c'}{r} = 0 \label{eq_mot_Q1_Q4_Q6_c}
  \end{align}
\end{subequations}
will reproduce the classical solution \eqref{BTZ_metric}. Notice that in the derivation
of \eqref{eq_mot_Q1_Q4_Q6} we chose
\begin{equation}
 \label{gauge_lapse}
  n(r) = -2\, \Lambda\,  r.
\end{equation}
The above gauge fixing of the lapse function is compatible with its expression \eqref{algebr_sol_n}
on the solution space and, in addition, also satisfies the Hamiltonian constraint.
Thus, our choice of the lapse \eqref{gauge_lapse} does not lead to inconsistencies.

Now, we will solve the system \eqref{eq_mot_Q1_Q4_Q6}. The last of the equations \eqref{eq_mot_Q1_Q4_Q6_c} leads to
\begin{equation}
 \label{sol_eq_mot_Q1_Q4_Q6_c}
   c(r) = c_1.
\end{equation}
The other two can also be directly integrated
\begin{subequations}
 \begin{align}
  & a(r) = \sqrt{\Lambda\, r^2 + c_2}, \label{sol_eq_mot_Q1_Q4_Q6_a} \\
  & b(r) = \sqrt{r^2 + c_3}. \label{sol_eq_mot_Q1_Q4_Q6_b}
 \end{align}
\end{subequations}
By choosing the constants appearing in the solutions above as $\{c_1, c_2, c_3\} =
\{\frac{J}{2}, -M, 0\}$, one exactly reproduces (as expected) the classical
solution \eqref{BTZ_metric}.

\subsection{The two dimensional subalgebra \{\texorpdfstring{$\widehat{Q}_1, \widehat{Q}_5$}{Q1, Q5}\}}
\label{sec:Q1_Q5_semi}

Let us now move to the solution \eqref{sol_Q1_Q5} of the two dimensional subalgebra
\eqref{2_subalg_a}. After substitution of the numerical values \eqref{kappas_0}, the
corresponding wave function looks like
\begin{equation}
 \label{wave_Q1_Q5}
  \Psi = a^{\ima\, M} b^{-\ima\, M} \left[c_1\, J_{-\ima\, M}\left(2\, a\, b\, \sqrt{\Lambda }\right) +
  c_2\, Y_{-\ima\, M}\left(2\, a\, b\, \sqrt{\Lambda }\right)\right].
\end{equation}
Because of the appearance of the Bessel functions in the above wave function we cannot
decisively conclude about the form of its amplitude and phase. To bring \eqref{wave_Q1_Q5}
into the form \eqref{wave} we have to study its behaviour for large and small values
of $r$ separately.

\subsubsection{Asymptotic behaviour for large values of r}
\label{sec:Q1_Q5_semi_asymp}

Notice that the asymptotic behaviour of the dynamical fields \eqref{BTZ_metric} is
$a(r) \sim r,\, b(r) \sim r,\, c(r) \sim \frac{J}{2}$; thus, the arguments of the
Bessel functions in \eqref{wave_Q1_Q5} in the limit of large $r$ are also large.
The behaviour of the Bessel functions for large values of its arguments is the
following \cite{Abramo}
\[
 J_\nu(z) \sim \sqrt{\frac{2}{\pi\, z}} \cos \left(\frac{1}{4} (-2\, \pi\, \nu + 4\, z - \pi)\right), \quad
 Y_\nu(z) \sim \sqrt{\frac{2}{\pi\, z}} \sin \left(\frac{1}{4} (-2\, \pi\, \nu + 4\, z - \pi )\right).
\]
Now, inserting the above asymptotic expressions of the Bessel functions into \eqref{wave_Q1_Q5} one gets
\be
\begin{split}
  \Psi \sim \frac{a^{-\frac{1}{2} + \ima\, M} b^{-\frac{1}{2} - \ima\, M}}{\sqrt{\pi}\, \Lambda^{1/4}}
  &\left[ c_1 \cos \left( \frac{1}{4} \left(8\, a\, b \sqrt{\Lambda} + 2\, \ima\, \pi M - \pi \right) \right) \right. \\
  &\left. +  c_2 \sin \left( \frac{1}{4} \left(8\, a\, b \sqrt{\Lambda } + 2\, \ima\, \pi M - \pi \right) \right) \right].
\end{split}
\ee
Notice that the simple choice of the constants $c_2 = \ima\, c_1$ brings the
expression in the square brackets into an exponential form; thus, we arrive at the
final form of the wave function in the asymptotic limit
\begin{equation}
 \label{wave_Q1_Q5_asymp}
  \Psi \sim c_3\, a^{-\frac{1}{2}}\, b^{-\frac{1}{2}}\, e^{\ima\,\left[2\, a\, b \sqrt{\Lambda } + M \ln(\frac{a}{b})\right]},
\end{equation}
where $c_3$ is a complex constant. From \eqref{wave_Q1_Q5_asymp} one can easily read
the amplitude and the phase, namely
\begin{equation}
 \label{ampl_phas_Q1_Q5_asymp}
  \Omega = a^{-\frac{1}{2}} b^{-\frac{1}{2}}, \qquad S = 2\, a\, b \sqrt{\Lambda } + M \ln\left(\frac{a}{b}\right),
\end{equation}
which satisfy the continuity equation but, unlike the previous case, give rise to a non-zero quantum potential
\begin{equation}
 \label{pot_Q1_Q5_asymp}
  \mathcal{V} = -\frac{1}{16\, a^2 b^2 \Lambda}.
\end{equation}
Thus, we expect that the solution to the semiclassical equations of motion
\begin{subequations}
 \label{eq_mot_Q1_Q5}
  \begin{align}
   & \frac{2\, a\, b\, b'}{r} + \frac{M}{a} + 2\, \sqrt{\Lambda}\, b = 0, \label{eq_mot_Q1_Q5_a} \\
   & \frac{2\, a\, a'\, b}{r} + 2 \sqrt{\Lambda}\, a - \frac{M}{b} = 0, \label{eq_mot_Q1_Q5_b} \\
   & \frac{c'}{r} = 0 \label{eq_mot_Q1_Q5_c}
  \end{align}
\end{subequations}
will differ from the classical solution \eqref{BTZ_metric}. Notice that here we chose the
same gauge \eqref{gauge_lapse} for the lapse function.

As before \eqref{eq_mot_Q1_Q5_c} can be readily integrated
\begin{equation}
 \label{sol_eq_mot_Q1_Q5_c}
   c(r) = c_1.
\end{equation}
An appropriate combination of the other two equations gives
\[
 a(r) = - \frac{\sqrt{\Lambda}\, r^2}{b},
\]
where an integration constant has been set to zero. Inserting the last expression into
\eqref{eq_mot_Q1_Q5_a} and integrating one obtains
\begin{equation}
 \label{sol_eq_mot_Q1_Q5_b}
  b(r) = c_2\, r\, e^{\frac{M}{4\, \Lambda\, r^2}},
\end{equation}
which leads immediately to
\begin{equation}
 \label{sol_eq_mot_Q1_Q5_a}
  a(r) = - \frac{r\, \sqrt{\Lambda}}{c_2}\, e^{-\frac{M}{4\, \Lambda\, r^2}}.
\end{equation}
Choosing appropriately the constants appearing in the solution above, i.e. $\{c_1, c_2\} =
\{\frac{J}{2}, 1\}$, we can exactly recover the classical solution \eqref{BTZ_metric} in
the limit $r \rightarrow \infty$: $a(r)^2 \sim r^2,\, b(r)^2 \sim r^2,\, c(r) \sim \frac{J}{2}$.
Interestingly enough, the above solution
\[
  a(r) = - r\, \sqrt{\Lambda}\, e^{-\frac{M}{4\, \Lambda\, r^2}}, \quad
  b(r) = r\, e^{\frac{M}{4\, \Lambda\, r^2}}, \quad c(r) = \frac{J}{2}
\]
does not exhibit an event horizon; a statement that can be easily verified after a quick
inspection of the ensuing line element
\begin{equation}
 \label{non_horizon_sol}
  ds^2 = - \Lambda\, r^2 e^{-\frac{M}{2\, \Lambda\, r^2}}\, dt^2 + \frac{4\, r^2}{J^2 + 4\, \Lambda\, r^4}\, dr^2 +\
  J\, dt\,d\phi + r^2 e^{\frac{M}{2 \Lambda  r^2}}\, d\phi^2.
\end{equation}
This is a space-time with much less symmetry than the classical geometry: it admits only the manifest Killing fields $\partial_t$ and $\partial_\phi$. Concerning the constants appearing in \eqref{non_horizon_sol}, all three of them are locally essential as it can seen by the fact that the relevant infinitesimal criterion \eqref{papcrit} (see \cite{gop}) admits no solution for any of $\Lambda$, $M$ and $J$. 

It is also worth mentioning that according to the form of the quantum potential \eqref{pot_Q1_Q5_asymp},
the quantum effects introduced by it must fade away as we approach $r \rightarrow \infty$,
while for smaller (but still large) values of $r$ the quantum effects must get stronger.
That is exactly the behaviour we observe for the above solution \eqref{non_horizon_sol},
where the quantum effects caused by the non-vanishing of the quantum potential must be
responsible for the disappearance of the event horizon.

\subsubsection{Behaviour close to the origin \texorpdfstring{$r = 0$}{r = 0}}
\label{sec:Q1_Q5_semi_origin}

Let us now see how is the wave function \eqref{wave_Q1_Q5} behaving near the origin
$r = 0$. In view of \eqref{BTZ_metric} the dynamical fields in the limit of small
$r$'s behave like $a(r) \sim M,\, b(r) \sim r,\, c(r) \sim \frac{J}{2}$; thus, likewise
the arguments of the Bessel functions in \eqref{wave_Q1_Q5} are also small close to
the origin. By inserting into \eqref{wave_Q1_Q5} the expression of the Bessel functions for small arguments \cite{Abramo}
\[
 J_\nu(z) \sim \frac{2^{-\nu}}{\Gamma (\nu + 1)}\, z^\nu, \quad
 Y_\nu(z) \sim -\frac{2^{\nu}\, \Gamma (\nu)}{\pi}\, z^{-\nu},
\]
where $ \Gamma$ is the gamma function, one obtains
\[
 \Psi \sim \frac{c_1\, \Lambda^{-\frac{\ima M}{2}}}{\Gamma (1 - \ima M)}\, b^{-2\, \ima M} -
 \frac{c_2\, \Lambda^{\frac{\ima M}{2}} \Gamma (-\ima M)}{\pi}\, a^{2\, \ima M},
\]
which after a re-definition of the constants reduces to
\[
  \Psi \sim c_3 \left(a^{2\, \ima M} + b^{-2\, \ima M}\right),
\]
where $c_3$ is a complex constant. Now, using the relation $y^{\ima x} = \cos(x \ln y) + \ima \sin(x \ln y)$
and some trigonometric identities, one can bring the above expression into the explicitly
exponential form
\begin{equation}
 \label{wave_Q1_Q5_origin}
  \Psi \sim 2\, c_3 \cos \left(M \ln (a\, b)\right)\, e^{\ima M \ln(\frac{a}{b})}.
\end{equation}
Comparing with \eqref{wave}, the amplitude and the phase of the wave function \eqref{wave_Q1_Q5_origin}
follow
\begin{equation}
 \label{ampl_phas_Q1_Q5_origin}
  \Omega = \cos \left(M \ln (a\, b)\right), \qquad S = M \ln\left( \frac{a}{b} \right),
\end{equation}
which as expected satisfies the continuity equation and leads to the non-vanishing
potential
\begin{equation}
 \label{pot_Q1_Q5_origin}
  \mathcal{V} = \frac{M^2}{4\, a^2\, b^2\, \Lambda }.
\end{equation}
The semiclassical equations of motion \eqref{semicl_eqs} read
\begin{subequations}
 \label{eq_mot_Q1_Q5_origin}
  \begin{align}
   & \frac{2\, a\, b\, b'}{r} + \frac{M}{a} = 0, \label{eq_mot_Q1_Q5_origin_a} \\
   & \frac{2\, a\, a'\, b}{r} - \frac{M}{b} = 0, \label{eq_mot_Q1_Q5_origin_b} \\
   & \frac{c'}{r} = 0 \label{eq_mot_Q1_Q5_origin_c},
  \end{align}
\end{subequations}
where the gauge \eqref{gauge_lapse} was again used. Integration of \eqref{eq_mot_Q1_Q5_origin_c}
yields
\begin{equation}
 \label{sol_eq_mot_Q1_Q5_origin_c}
   c(r) = c_1.
\end{equation}
An appropriate combination of the other two equations leads to the condition
\[
 a(r) = \frac{1}{b(r)},
\]
where an integration constant was set to unity as it does not change the geometry of the resulting line element. Inserting the above condition into
\eqref{eq_mot_Q1_Q5_origin_b}, one can solve for $b(r)$ and arrive (after setting the
integration constant to unity) at
\begin{equation}
 \label{sol_eq_mot_Q1_Q5_origin_b}
   b(r) = e^{-\frac{1}{4}\, M r^2},
\end{equation}
which readily leads to the form of $a(r)$:
\begin{equation}
 \label{sol_eq_mot_Q1_Q5_origin_a}
   a(r) = e^{\frac{1}{4}\, M r^2}.
\end{equation}
Setting $c_1 = \frac{J}{2}$ in \eqref{sol_eq_mot_Q1_Q5_origin_c}, the resulting line
element for the solution \eqref{sol_eq_mot_Q1_Q5_origin_c}-\eqref{sol_eq_mot_Q1_Q5_origin_a}
reads
\begin{equation}
 \label{non_central_sing_sol}
  ds^2 = -e^{\frac{M r^2}{2}}\, dt^2 + \frac{4\, r^2}{J^2+4}\, dr^2 + J\, dt\,d\phi + e^{-\frac{M r^2}{2}}\, d\phi^2.
\end{equation}
This line element represents a homogeneous space-time, admitting the three Killing fields
$\partial_t$, $\partial_\phi$ and $-M\partial_t+\frac{2}{r}\partial_r+M\partial_\phi$. Notice that,
since there is no classical curvature singularity, we would expect the semiclassical geometry to share
the same property; indeed, for the above metric, there is no central singularity at $r=0$. An interesting
observation is that, as earlier mentioned, both $M$ and $J$ appearing in the metric are now also locally
essential constants; the Ricci scalar is $R=-\frac{M^2}{2}$ and the Kretchmann scalar is
$R^{ijkl}R_{ijkl}=-\frac{1}{4} \left(2 J^2-3\right) M^4$. Another property is that the Killing horizon
of the classical black hole has disappeared.

\subsection{The two dimensional subalgebra \{\texorpdfstring{$\widehat{Q}_2, \widehat{Q}_4$}{Q2, Q4}\}}
\label{sec:Q2_Q4_semi_origin}

We will continue our semiclassical study with the wave function \eqref{sol_Q2_Q4} of
the two dimensional subalgebra \eqref{2_subalg_b}. After substituting the numerical
values \eqref{kappas_0} into \eqref{sol_Q2_Q4} the wave function reduces to
\[
 \Psi = \frac{c_0}{a}\, e^{\ima \left(-\frac{\Lambda  (J-2 c)^2}{4\, a^2} - a^2 - b^2 \Lambda \right)},
\]
which is already in exponential form; therefore, one can define
\begin{equation}
 \label{ampl_phas_Q2_Q4}
  \Omega = \frac{c_0}{a}, \qquad S = -\frac{\Lambda  (J-2 c)^2}{4\, a^2} - a^2 - b^2 \Lambda.
\end{equation}
The above amplitude and phase satisfy the continuity equation, result to a vanishing
quantum potential $\mathcal{V} = 0$, and lead to the semiclassical equations of motion
\begin{subequations}
 \label{eq_mot_Q2_Q4}
  \begin{align}
   & \frac{2\, a\, b\, b'}{r} - \frac{2\, J\, \Lambda\, c}{a^3} + \frac{2\, \Lambda\, c^2}{a^3} +
     \frac{J^2 \Lambda}{2\, a^3} - 2\, a = 0, \label{eq_mot_Q2_Q4_a} \\
   & \frac{2\, a\, a'\, b}{r} - 2\, \Lambda\, b = 0, \label{eq_mot_Q2_Q4_b} \\
   & \frac{2\, \Lambda\, c}{a^2} - \frac{J\, \Lambda}{a^2} - \frac{c'}{r} = 0 \label{eq_mot_Q2_Q4_c},
  \end{align}
\end{subequations}
where the condition \eqref{gauge_lapse} was again used. Equation \eqref{eq_mot_Q2_Q4_b}
can be directly integrated
\begin{equation}
 \label{sol_eq_mot_Q2_Q4_a}
   a(r) = \sqrt{\Lambda\, r^2 + c_1}\,.
\end{equation}
Now, inserting \eqref{sol_eq_mot_Q2_Q4_a} into \eqref{eq_mot_Q2_Q4_c} one gets
\begin{equation}
 \label{sol_eq_mot_Q2_Q4_c}
   c(r) = \frac{J}{2},
\end{equation}
where an integration constant was set to zero. Finally, by substituting the expressions
\eqref{sol_eq_mot_Q2_Q4_a}-\eqref{sol_eq_mot_Q2_Q4_c} into \eqref{eq_mot_Q2_Q4_b}
and integrating, one arrives at the solution
\begin{equation}
 \label{sol_eq_mot_Q2_Q4_b}
   b(r) = \sqrt{r^2 + c_2}\,.
\end{equation}
By choosing the constants that appear in the derivation above as $\{c_1, c_2\} = \{-M, 0\}$,
the solution \eqref{sol_eq_mot_Q2_Q4_a}-\eqref{sol_eq_mot_Q2_Q4_b}, as expected because
of the vanishing of the quantum potential, reproduces the classical solution \eqref{BTZ_metric}.

\subsection{The two dimensional subalgebra \{\texorpdfstring{$\widehat{Q}_3, \widehat{Q}_6$}{Q3, Q6}\}}
\label{sec:Q3_Q6_semi_origin}

Let us conclude the semiclassical analysis with the subalgebra \eqref{2_subalg_c}.
The corresponding quantum solution \eqref{sol_Q3_Q6}, after the substitution of
the numerical values \eqref{kappas_0}, reduces to the wave function
\[
 \Psi = \frac{c_0}{b}\,e^{\ima \left( -a^2 - b^2 \Lambda - \frac{c^2}{b^2} + \frac{c\, J}{b^2} - \frac{J^2}{4\, b^2} \right)},
\]
which is already in the form of \eqref{wave}; thus, the amplitude and the phase
readily follow
\begin{equation}
 \label{ampl_phas_Q3_Q6}
  \Omega = \frac{c_0}{b}, \qquad S = -a^2 - b^2 \Lambda - \frac{c^2}{b^2} + \frac{c\, J}{b^2} - \frac{J^2}{4\, b^2}.
\end{equation}
In view of \eqref{ampl_phas_Q3_Q6} the continuity equation is identically satisfied,
the quantum potential \eqref{quant_pot} is trivial, and the equations of motion
\eqref{semicl_eqs} in the gauge \eqref{gauge_lapse} read
\begin{subequations}
 \label{eq_mot_Q3_Q6}
  \begin{align}
   & \frac{2\, a\, b\, b'}{r} - 2\, a = 0, \label{eq_mot_Q3_Q6_a} \\
   & \frac{2\, a\, a'\, b}{r} - \frac{2\, J\, c}{b^3} + \frac{2\, c^2}{b^3} + \frac{J^2}{2\, b^3} -
     2\, \Lambda\,  b = 0, \label{eq_mot_Q3_Q6_b} \\
   & \frac{2\, c}{b^2} - \frac{J}{b^2} - \frac{c'}{r} = 0 \label{eq_mot_Q3_Q6_c}.
  \end{align}
\end{subequations}
Obviously, equation \eqref{eq_mot_Q3_Q6_a} can be directly integrated, yielding
\begin{equation}
 \label{sol_eq_mot_Q3_Q6_b}
   b(r) = \sqrt{r^2 + c_1}\,.
\end{equation}
Insertion of the above result into \eqref{eq_mot_Q3_Q6_c} and integrating one obtains
\begin{equation}
 \label{sol_eq_mot_Q3_Q6_c}
   c(r) = \frac{J}{2},
\end{equation}
where, as before, an integration constant was set to zero. Finally, taking into
account the expressions for $b(r)$ and $c(r)$, the last remaining equation
\eqref{eq_mot_Q3_Q6_b} of the above system admits the solution
\begin{equation}
 \label{sol_eq_mot_Q3_Q6_a}
   a(r) = \sqrt{\Lambda\, r^2 + c_2}\,.
\end{equation}
An appropriate choice of the integration constants appearing above, i.e. $\{c_1, c_2\} =
\{0, -M\}$, identifies the solution \eqref{sol_eq_mot_Q3_Q6_b}-\eqref{sol_eq_mot_Q3_Q6_a}
with the classical solution space given in \eqref{BTZ_metric}.

%%%%%%%%%%%%%%%%%%%%%%%%%%%%%%%%%%%%%%%%%%%%%%%%%%%%%%%%%%%%%%%%%%%%%%%%%%%%%%%%%%%%%%%%

\section{Discussion}
\label{discussion}

In the present work the method of canonical quantization has been implemented for the BTZ geometry.
The procedure begins at the classical level with the construction of a ``proper" mini-superspace
description. Accordingly, a Lagrangian function that describes a system dynamically equivalent
to that of the Einstein field equations of 2+1 gravity in the presence of a cosmological constant
must be derived.

The resulting Lagrangian is of course singular, thus one has to take into account the modified infinitesimal
criterion \eqref{crit} in order to acquire all the variational symmetries of the action. To simplify
the procedure, we chose to work in the constant potential parametrization. In this parametrization
the $\xi$'s of \eqref{crit} become Killing fields of the corresponding mini-supermetric, which is
three dimensional, flat and exhibits a homothecy. The six Killing fields and the homothetic field
are linked to six autonomous \eqref{int_motion} and a rehonomic \eqref{homo_quant} integrals of motion,
respectively. In phase space we use these seven conserved quantities, not only to derive the BTZ solution
algebraically, but also to read off the two Casimir invariants of the six dimensional algebra spanned
by the $Q_I$'s.

The adoption of the constant potential parametrization becomes imperative at the quantum level. There
the non-identically zero Casimir invariant is manifested as the kinetic part of the Hamiltonian. Hence,
the Wheeler-DeWitt operator becomes compatible with any choice of the eigen-operators $\widehat{Q}_I$
that satisfy the integrability conditions \eqref{selec_rule}. There exist four independent maximal Abelian
subalgebras of the $\widehat{Q}_I$'s that lead to an equal number of different wave functions that also
satisfy the constraint $\widehat{\mathcal{H}}\Psi=0$, one for each set of ``measurable" quantities.

Using Bohm's approximation for the derivation of quantum trajectories, one can assign to each wave
function a corresponding semiclassical geometry. Our analysis shows that for the subalgebras \eqref{3_subalg},
\eqref{2_subalg_b}, and \eqref{2_subalg_c} the corresponding geometries are identical to the classical
one due to the fact that the quantum potential is zero. A possible explanation of this result follows
from the particular lapse parametrization we are using here, which identifies the kinetic part of the
quantum quadratic constraint with the Casimir invariant of the algebra of the charges (see the relevant
section in \cite{tchris3} for a complete discussion). However, the simultaneous measurement of $Q_1$
and $Q_5$ leads to quantum corrections at both limits $r\rightarrow +\infty$ and $r\rightarrow 0$:
In the asymptotic limit, one obtains a semiclassical geometry that is asymptotically AdS ($R=-6\Lambda -\frac{M^2}{2\Lambda r^4}$),
which is in accordance with the classical geometry, where $R=-6\Lambda$. The existence of a curvature
singularity at $r=0$, in this semiclassical approximation, is not alarming since the region of validity
of the solution rests in the range of large values of $r$. In the vicinity of the origin, $r\rightarrow 0$,
the semiclassical geometry \eqref{non_central_sing_sol} is obtained. A quick inspection of the line
element \eqref{non_central_sing_sol} entails that the aforementioned ``macroscopic singularity" does
not exist anymore. Moreover, this homogeneous semiclassical space-time is characterized by two essential
constants, i.e. $M$ and $J$. The latter can be interpreted as an indication that at small scales the
contribution of the mass and of the angular momentum exceeds by far the corresponding contribution of
the cosmological constant $\Lambda$.

\begin{appendices}

\section{Local elimination of constants} \label{appA}

In \cite{gop}, the following infinitesimal criterion that determines whether a particular constant (say $\lambda$)
that appears in a space-time metric $g_{\mu\nu}$ is not essential has been presented:
\be \label{papcrit}
\pound_\xi g_{\mu\nu}+\partial_\lambda g_{\mu\nu}=0.
\ee
One can interpret \eqref{papcrit} as stating that the change in form of the given metric, induced by the vector $\xi$
(through moving along its integral curves), is counterbalanced by the change induced on it through an infinitesimal
displacement of $\lambda$. In this spirit, whenever this set of coupled first order PDEs admits a solution $\xi$,
the coordinate transformation linked to its integral lines can be used for the elimination of the corresponding
constant $\lambda$.

For the case of the BTZ metric \eqref{BTZ_metric} the corresponding relations are
\begin{subequations} \label{speccrit}
\begin{align}
&\pound_{\xi_M} g_{\mu\nu}+\partial_M g_{\mu\nu} =0 \\
&\pound_{\xi_J} g_{\mu\nu}+\partial_J g_{\mu\nu} =0
\end{align}
\end{subequations}
and can be readily solved to give
\begin{subequations} \label{elimfields}
\begin{align}\label{elimfieldsM}
\xi_M & = \frac{M t-\phi  J}{2 J^2 \Lambda -2 M^2}\partial_t +\frac{J^2-2 M r^2}{4 J^2 \Lambda  r-4 M^2 r}\partial_r +\frac{\phi  M-J \Lambda  t}{2 J^2 \Lambda -2 M^2}\partial_\phi \\ \label{elimfieldsJ}
\xi_J & = \frac{\phi  M-J \Lambda  t}{2 J^2 \Lambda -2 M^2}\partial_t -\frac{J M-2 J \Lambda  r^2}{4 J^2 \Lambda  r-4 M^2 r}\partial_r +\frac{\Lambda  (M t-\phi  J)}{2 J^2 \Lambda -2 M^2}\partial_\phi .
\end{align}
\end{subequations}
These fields have a vanishing Lie bracket, if viewed as fields in a space spanned by $t$, $r$, $\phi$, $M$ and $J$,
by considering the trivial addition of $\partial_M$ and $\partial_J$ in \eqref{elimfieldsM} and \eqref{elimfieldsJ},
respectively. We can therefore try to find their common invariant functions, the number of which is expected (in
view of the linear independence of $\xi_M$ and $\xi_J$) to be $5-2=3$. Indeed a straightforward calculation reveals
that an appropriate set is
\begin{subequations} \label{invfunc}
\begin{align}
f_1(t,r,\phi,M,J) &=\sqrt{J \sqrt{\Lambda }+M} \left(\phi +\sqrt{\Lambda } t\right) \\
f_2(t,r,\phi,M,J) &=\frac{M-2 \Lambda  r^2}{2 \Lambda  \sqrt{M^2-J^2 \Lambda }} \\
f_3(t,r,\phi,M,J) &=\sqrt{\Lambda  \left(M-J \sqrt{\Lambda }\right)} \left(\phi -\sqrt{\Lambda } t\right).
\end{align}
\end{subequations}
One can easily check that \eqref{invfunc} satisfy $\xi_M^\mu \partial_\mu f_i+\partial_M f_i=0$ and $\xi_J^\mu \partial_\mu f_i+\partial_J f_i=0$ for $i=1,2,3$.
By considering the $f_i$'s as the new set of variables, say $\tau=f_1$, $\rho=f_2$ and $\theta=f_3$, we can calculate
the transformation from the old variables to the new $(t,r,\phi)\longrightarrow (\tau,\rho,\theta)$:
\begin{subequations} \label{elimtr}
\begin{align}
t &= \frac{\tau }{2 \sqrt{\Lambda } \sqrt{J \sqrt{\Lambda }+M}}- \frac{\theta }{2 \Lambda  \sqrt{M-J \sqrt{\Lambda }}} \\
r &= \sqrt{\frac{M}{2 \Lambda }-\rho  \sqrt{M^2-J^2 \Lambda }} \\
\phi &= \frac{1}{2} \left(\frac{\tau }{\sqrt{J \sqrt{\Lambda }+M}}+\frac{\theta }{\sqrt{\Lambda  \left(M-J \sqrt{\Lambda }\right)}}\right)
\end{align}
\end{subequations}

The Jacobian matrix of the transformation is
\be
J^\mu_\nu=\left(
\begin{array}{ccc}
 \frac{1}{2 \sqrt{\sqrt{\Lambda } J+M} \sqrt{\Lambda }} & 0 & -\frac{1}{2 \sqrt{\Lambda } \sqrt{\left(M-J \sqrt{\Lambda }\right) \Lambda }} \\
 0 & -\frac{\sqrt{M^2-J^2 \Lambda }}{\sqrt{\frac{2 M}{\Lambda }-4 \sqrt{M^2-J^2 \Lambda } \rho }} & 0 \\
 \frac{1}{2 \sqrt{\sqrt{\Lambda } J+M}} & 0 & \frac{1}{2 \sqrt{\left(M-J \sqrt{\Lambda }\right) \Lambda }} \\
\end{array}
\right)
\ee
having a non-zero determinant $J=-\frac{1}{2 \sqrt{2} \sqrt{\Lambda  \left(M-2 \Lambda  \rho  \sqrt{M^2-J^2 \Lambda }\right)}}$.
The metric $g_{\kappa\lambda}$ in the new variables reads
\be
\tilde{g}_{\mu\nu}:=g_{\kappa\lambda}J^\kappa_\mu J^\lambda_\nu = \left(
\begin{array}{ccc}
 \frac{1}{4 \Lambda } & 0 & -\frac{\rho }{2 \sqrt{\Lambda }} \\
 0 & \frac{\Lambda }{4 \Lambda ^2 \rho ^2-1} & 0 \\
 -\frac{\rho }{2 \sqrt{\Lambda }} & 0 & \frac{1}{4 \Lambda ^2} \\
\end{array}
\right)
\ee
and is thus brought into a form free of $M$ and $J$, with $\Lambda$ appearing as the sole essential constant.
Needless to say that, either in the original coordinates or in the final, an equation analogous to \eqref{speccrit}
concerning $\Lambda$ admits no solution. The same applies for $M$, $J$ appearing in the semiclassical solution
\eqref{non_central_sing_sol}.

\end{appendices}
%%%%%%%%%%%%%%%%%%%%%%%%%%%%%%%%%%%%%%%%%%%%%%%%%%%%%%%%%%%%%%%%%%%%%%%%%%%%%%%%%%%%%%%%

\end{document}